\documentclass[twocolumn]{aastex631}

\newcommand{\teff}{$T_{\mathrm{eff}}$}

\newcommand{\rearth}{$\mathrm{R_\oplus}$}
\newcommand{\fearth}{$\mathrm{F_\oplus}$}
\newcommand{\kep}{{\it Kepler}}
\newcommand{\ktwo}{{\it K2}}
\newcommand{\tess}{{\it TESS}}
\newcommand{\gaia}{{\it Gaia}}
\newcommand{\rprstar}{$\frac{R_{\mathrm{p}}}{R_\star}$}
\newcommand{\msun}{$\mathrm{M_\odot}$}
\newcommand{\rsun}{$\mathrm{R_\odot}$}

\newcommand{\calcalpha}{0.069$^{+0.019}_{-0.023}$}
\newcommand{\calcbeta}{$-$0.046$^{+0.125}_{-0.117}$}

\usepackage{hyperref}
\usepackage{amsmath}

\makeatletter
\newcommand{\labeltext}[2]{%
  \@bsphack
  \csname phantomsection\endcsname 
  \def\@currentlabel{#1}{\label{#2}}%
  \@esphack
}

\shorttitle{Analyzing the 3D Exoplanet Radius Gap}
\shortauthors{Berger et al.}

\begin{document}

\title{Evidence that Core-Powered Mass-Loss Dominates Over Photoevaporation in Shaping the Kepler Radius Valley}

\correspondingauthor{Travis Berger}
\email{taberger@hawaii.edu}

\author[0000-0002-2580-3614]{Travis A. Berger}
\altaffiliation{NASA Postdoctoral Program Fellow}
\affiliation{Exoplanets and Stellar Astrophysics Laboratory, Code 667, NASA Goddard Space Flight Center, Greenbelt, MD, 20771, USA}

\author[0000-0001-5347-7062]{Joshua E. Schlieder}
\affiliation{Exoplanets and Stellar Astrophysics Laboratory, Code 667, NASA Goddard Space Flight Center, Greenbelt, MD, 20771, USA}

\author[0000-0001-8832-4488]{Daniel Huber}
\affiliation{Institute for Astronomy, University of Hawai`i, 2680 Woodlawn Drive, Honolulu, HI 96822, USA}

\author[0000-0001-7139-2724]{Thomas Barclay}
\affiliation{Exoplanets and Stellar Astrophysics Laboratory, Code 667, NASA Goddard Space Flight Center, Greenbelt, MD, 20771, USA}
\affiliation{University of Maryland, Baltimore County, 1000 Hilltop Cir, Baltimore, MD 21250, USA}

\begin{abstract}
The dearth of planets with sizes around 1.8 $\mathrm{R_\oplus}$ is a key demographic feature discovered by the $Kepler$ mission. Two theories have emerged as potential explanations for this valley:  photoevaporation and core-powered mass-loss. However, \citet{Rogers2021b} shows that differentiating between the two theories is possible using the three-dimensional parameter space of planet radius, incident flux, and stellar mass. We use homogeneously-derived stellar and planetary parameters to measure the $Kepler$ exoplanet radius gap in this three-dimensional space. We compute the slope of the gap as a function of incident flux at constant stellar mass ($\alpha$ $\equiv$ $\left(\partial \log R_{\mathrm{gap}} / \partial \log S \right)_{M_\star}$) and the slope of the gap as a function of stellar mass at constant incident flux ($\beta$ $\equiv$ $\left(\partial \log R_{\mathrm{gap}} / \partial \log M_\star \right)_{S}$) and find $\alpha$ = \calcalpha\ and $\beta$ = \calcbeta. Given that \citet{Rogers2021b} shows that core-powered mass-loss predicts $\alpha$ $\approx$ 0.08 and $\beta$ $\approx$ 0.00 while photoevaporation predicts $\alpha$ $\approx$ 0.12 and $\beta$ $\approx$ --0.17, our measurements are more consistent with core-powered mass-loss than photoevaporation. However, we caution that different gap-determination methods can produce systematic offsets in both $\alpha$ and $\beta$; therefore, we motivate a comprehensive re-analysis of $Kepler$ light curves with modern, updated priors on eccentricity and mean stellar density to improve both the accuracy and precision of planet radii and subsequent measurements of the gap.
\end{abstract}

\section{Introduction}

One of the key exoplanet demographic results from the \kep\ Mission \citep{Borucki2010} was the discovery of a gap or dearth of planets with radii between 1.3 and 2.6 \rearth\ \citep{owen13,Fulton2017} at orbital periods $<$100 days. More precise stellar radii provided by the California-\kep\ Survey \citep{Petigura2017,Johnson2017} enabled this unambiguous discovery, and successive leveraging of \gaia\ parallaxes \citep{Brown2018,Lindegren2018,Berger2018c,Fulton2018} further cemented the gap as a demographic feature of \kep\ exoplanets. Moreover, it appears that the gap occurs in the \ktwo\ planet sample \citep{Hardegree2020,Zink2021} and may be present in other short-period exoplanet populations throughout our Galaxy.

Subsequent investigations have revealed how the gap varies as a function of orbital period \citep{Fulton2018,Berger2018c,VanEylen2018,Cloutier2020b,Petigura2022,Ho2023}, incident flux \citep{Fulton2018,Berger2018c,Berger2020b,Petigura2022,Ho2023}, stellar mass \citep{Fulton2018,Wu2019,Berger2020b,Cloutier2020b,Petigura2022,Ho2023}, stellar metallicity \citep{Petigura2018,Owen2018}, and stellar age \citep{Berger2018,Berger2020b,David2021,Sandoval2021,Petigura2022,Ho2023}. A number of theories have been introduced to explain the existence of the exoplanet radius gap, including planetesimal impacts \citep{Schlichting2015} to gas-poor formation \citep{Lee2014,Lee2016,Lee2019} to photoevaporation \citep{owen13,owen16,Owen2017,Owen2018,Wu2019,Rogers2021} to core-powered mass-loss \citep{Ginzburg2016,Ginzburg2018,Gupta2019,Gupta2020,Misener2021}. Photoevaporation and core-powered mass-loss have emerged as the most likely candidates for planets orbiting solar mass stars \citep[for low mass stars, see][]{Cloutier2020b}.

Photoevaporation requires incident extreme ultraviolet radiation to drive atmospheric escape, while core-powered mass-loss relies on both the core luminosity of a planet following its formation and the incident bolometric flux to strip planet atmospheres. According to \citet{Rogers2021b}, it is possible to discriminate between the two theories in the three-dimensional parameter space of planet radius, incident flux, and stellar mass. Here we aim to measure the 3D planet radius gap to differentiate between core-powered mass-loss and photoevaporation as the dominant mechanism sculpting the short period planet population.

\section{Stellar and Planet Samples}
\label{sec:samples}

\begin{figure*}
    \centering
    \resizebox{\hsize}{!}{\includegraphics{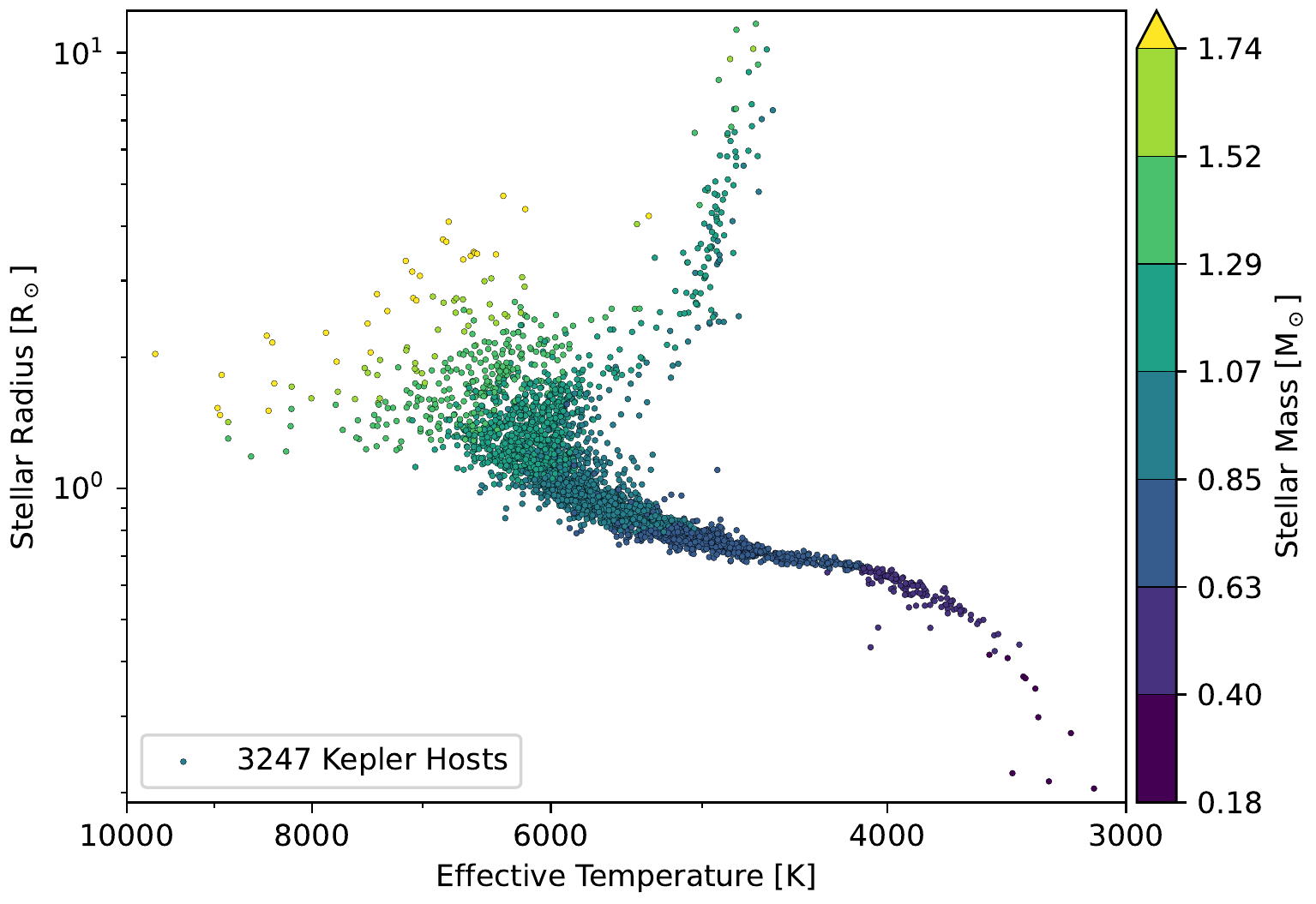}}
    \caption{Stellar radius versus effective temperature for the \kep\ host sample from \ref{B23}, which used \gaia\ DR3 photometry, parallaxes, spectrophotometric metallicities, extinction maps, and isochrone fitting with \texttt{isoclassify} to determine fundamental stellar parameters. Stars are colored by their mass.}
    \label{fig:HostHR}
\end{figure*}

\begin{figure*}
    \centering
    \resizebox{\hsize}{!}{\includegraphics{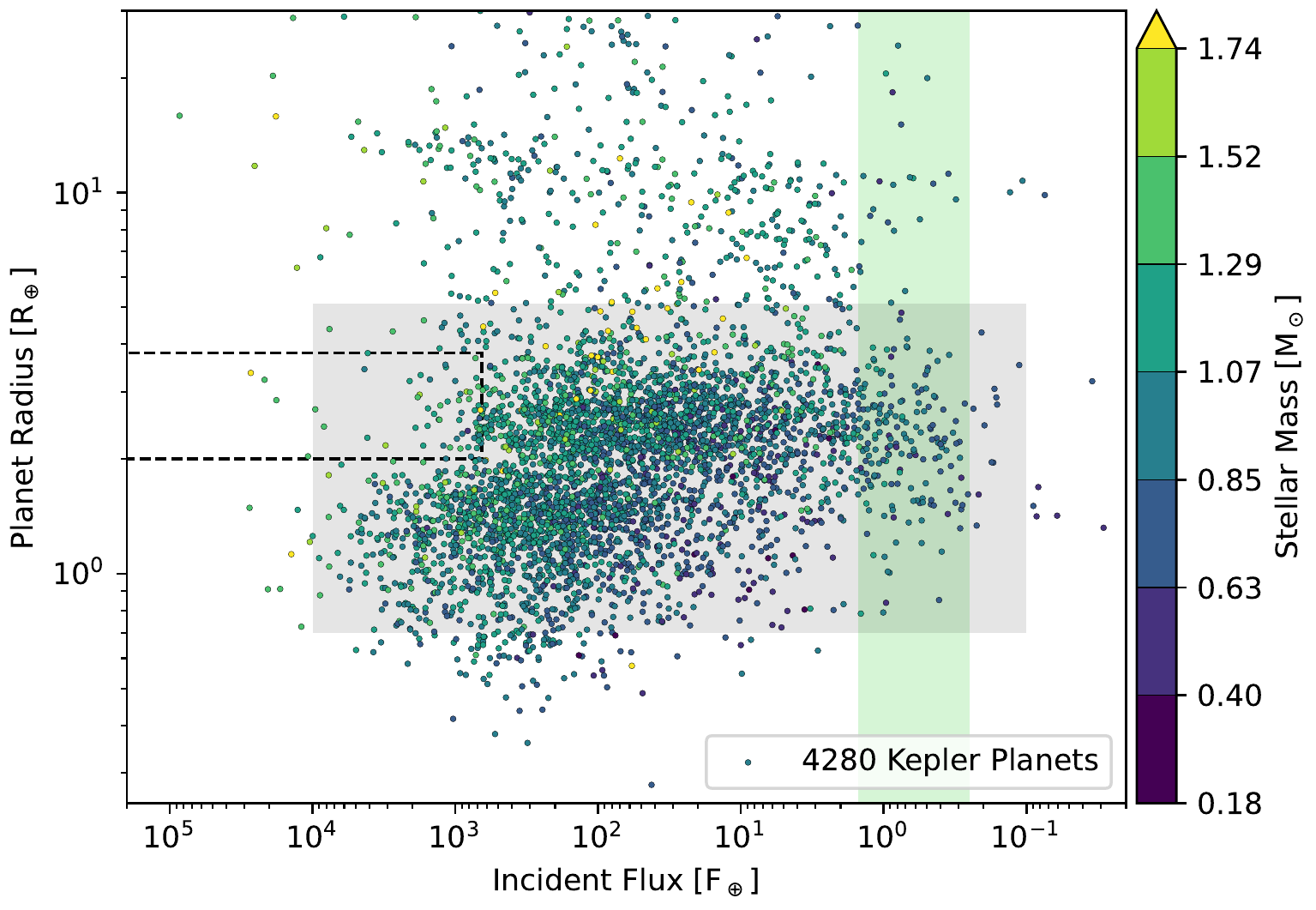}}
    \caption{Planet radius versus incident flux for the \kep\ planet sample from \ref{B23}. Planets are colored according to the stellar mass of their host star, and those within the grey shaded region (0.7 \rearth\ $\leq$ $R_{\mathrm{p}}$ $\leq$ 5.1 \rearth; 0.1 \fearth\ $\leq$ $F_{\mathrm{p}}$ $\leq$ 10$^4$ \fearth) will be included in the 3D planet radius gap measurement. We have also highlighted the hot sub-Neptunian desert within the dashed-box \citep[2 \rearth\ $\leq$ $R_{\mathrm{p}}$ $\leq$ 3.8 \rearth; $F_{\mathrm{p}}$ $>$ 650 \fearth,][]{Lundkvist2016} and the habitable zone as the green shaded region \citep[0.25 \fearth\ $\leq$ $F_{\mathrm{p}}$ $\leq$ 1.50 \fearth,][]{Kane2016}.}
    \label{fig:PlanetPradIncFlux}
\end{figure*}

We take the stellar parameters from Berger et al. (2023)\labeltext{B23}{B23} (hereafter \ref{B23}), which used \gaia\ DR3 photometry, parallaxes, positions, and spectrophotometric metallicities \citep{gaia1,GaiaEDR3,Lindegren2021a,Lindegren2021b,Riello2021,GaiaDR3,Babusiaux2022,Creevey2022,Fouesneau2022,Andrae2022}, and \texttt{isoclassify} \citep{Huber2017,Berger2020a} in combination with the all-sky \texttt{mwdust} map \citep{Bovy2016,Drimmel2003,schlafly14,Green2019} to derive stellar parameters from a custom-interpolated PARSEC model grid \citep{bressan12} with YBC synthetic photometry \citep{Chen2019}. As described in \ref{B23}, we fixed saturated photometry using the prescriptions in the appendix of \citet{Riello2021}, fixed parallax zero points using the prescriptions of \citet{Lindegren2021b}, and fixed the spectrophotometric metallicities using polynomial prescriptions (see \ref{B23}) based on the California-\kep\ Survey's \citep[CKS,][]{Petigura2017} metallicities for overlapping stellar samples.

These new stellar parameters supersede those of \citet{Berger2020a} because we use \gaia\ DR3 photometry, which is more precise and homogeneous than KIC photometry \citep{brown11}. In addition, we use calibrated \gaia\ metallicities, which are available for more \kep\ hosts and homogeneously derived unlike the heterogeneous combination of LAMOST \citep{LAMOSTDR5}, APOGEE \citep{APOGEEDR14}, and CKS \citep{Petigura2017} metallicities. We also improved our M-dwarf stellar parameters by combining YBC synthetic photometry \citep{Chen2019} and PARSEC models \citep{bressan12} with the \citet{mann15,Mann2019} empirical relations.

In Figure \ref{fig:HostHR}, we plot the \kep\ \citep{kepcumulative} host star radii as a function of effective temperature, colored by stellar mass. The \kep\ sample prioritized FGK dwarfs \citep{batalha10,Wolniewicz2021}, as the vast majority of \kep\ host stars have \teff\ from 4000--7000 K with corresponding masses from $\sim$0.6--1.3 \msun. Consequently, there are few hosts with masses $>$ 1.74 \msun\ and $<$ 0.4 \msun\ or stellar radii $>$ 3 \rsun. We removed KOI 5635 (KIC 9178894) from further analysis, as it is an outlier at \teff\ $\approx$ 12000 K with a mass of 3.4 \msun.

In Figure \ref{fig:PlanetPradIncFlux}, we plot the \kep\ confirmed and candidate planet radii as a function of incident flux, also colored by host star mass. We find that most \kep\ planets are smaller than 5 \rearth, indicating their high occurrence rates throughout the Galaxy \citep{Howard2010,Howard2012,Fressin2013,petigura13,petigura13b,dressing13,Hardegree2019}. In this paper, we focus on the 3702 planets within the grey box to isolate the populations of super-Earths below $\sim$2 \rearth\ and sub-Neptunes above $\sim$2 \rearth\ that are separated by the planet radius valley. In addition to the valley, we see, as in \citet{Berger2018c,Berger2020b}, a hot-Jupiter inflation trend \citep{miller11,Thorngren2016,Grunblatt2017}, planets within the hot sub-Neptunian desert \citep[dashed box,][]{Lundkvist2016}, and planets in the habitable zone \citep[green shaded region,][]{Kane2016}.

In order to preserve homogeneity in the derived planet parameters, we do not use \ktwo\ and \tess\ exoplanets, as \kep, \ktwo, and \tess\ used different transit-fitting pipelines. We chose \kep\ because it has superior precision relative to \ktwo\ and \tess\ and $>$4000 planets.

\section{Measuring the 3D Planet Radius Gap}
\label{sec:methods}

\begin{figure*}
    \centering
    \resizebox{0.49\hsize}{!}{\includegraphics{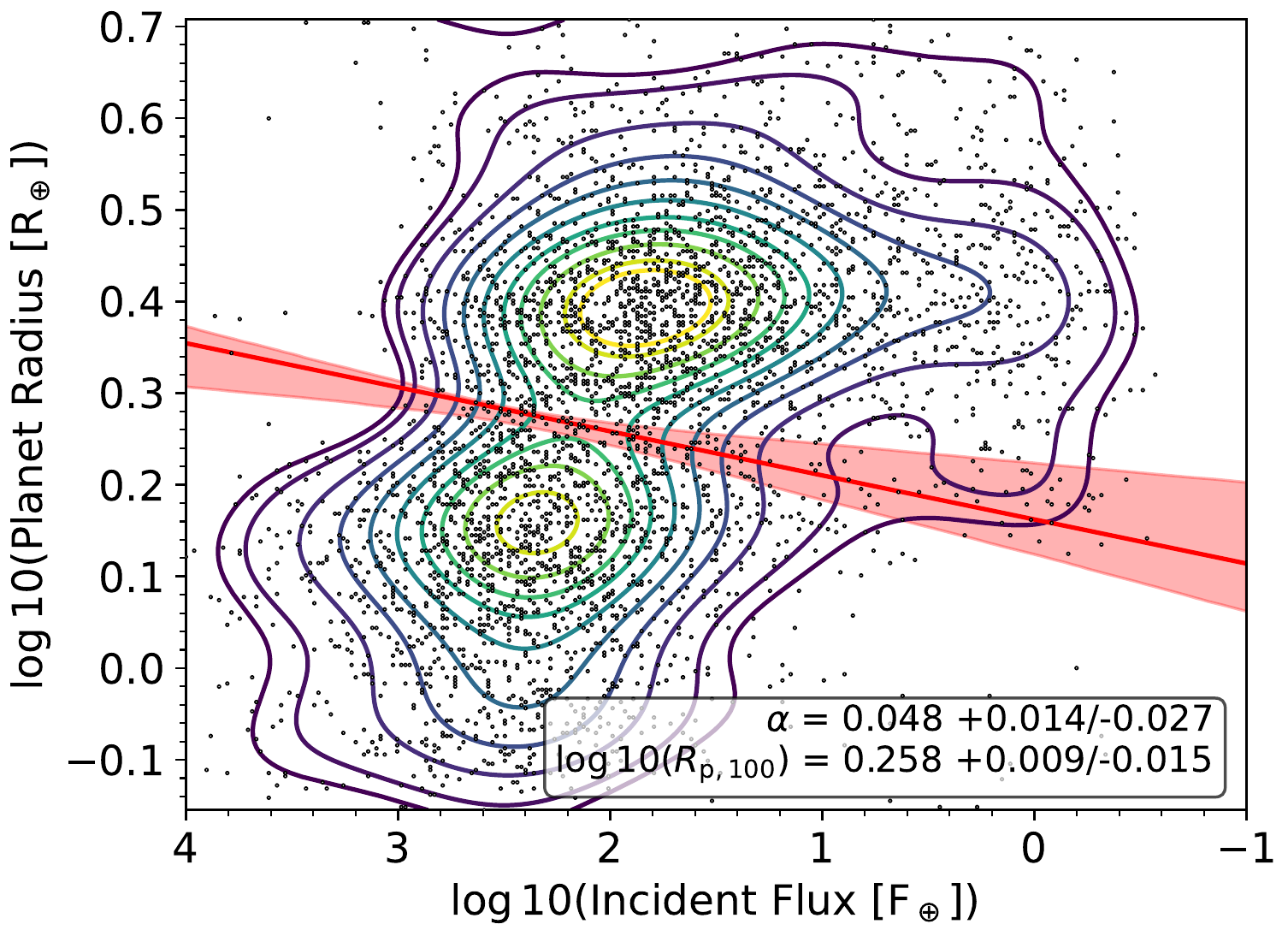}}
    \resizebox{0.49\hsize}{!}{\includegraphics{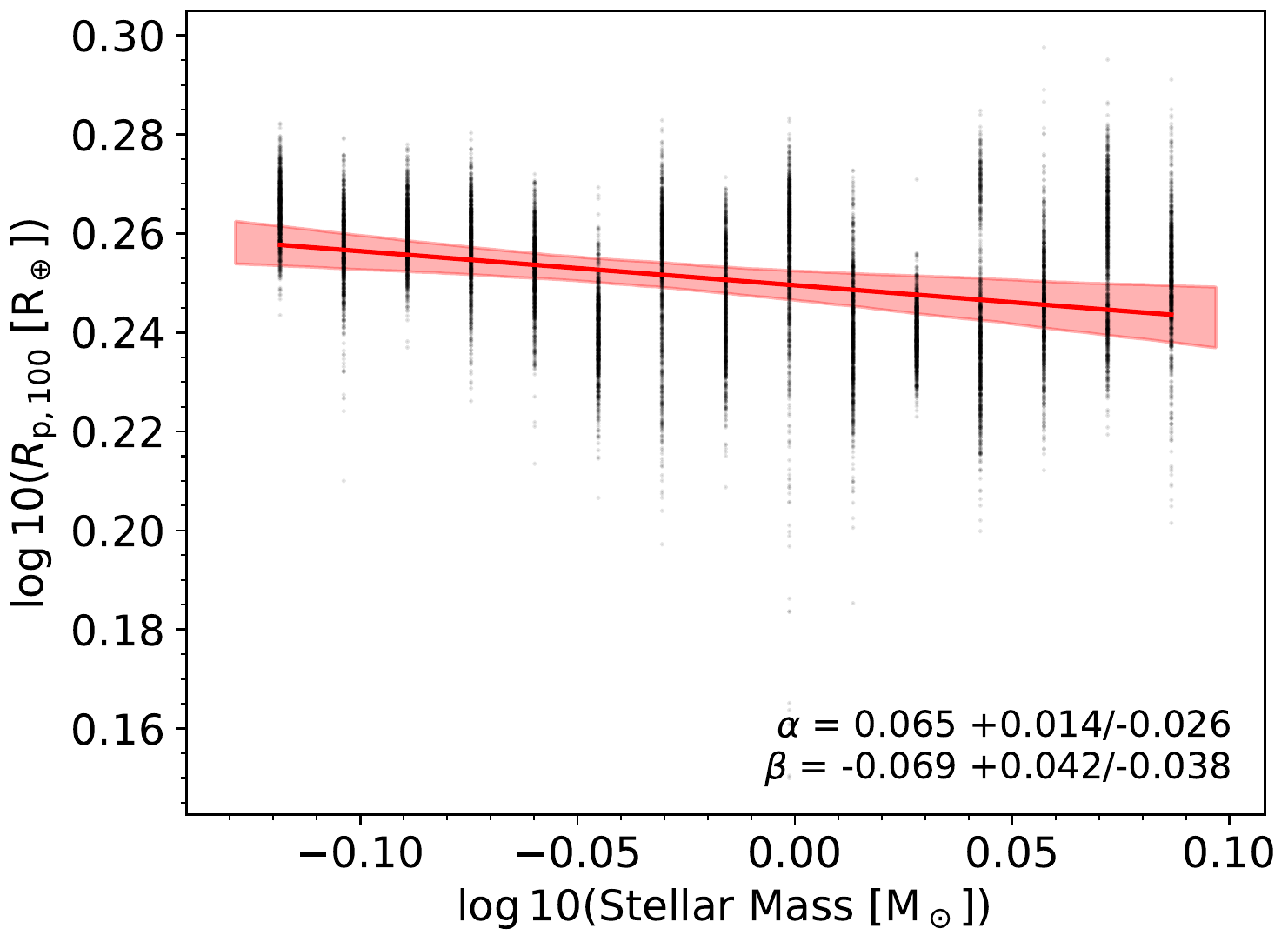}}
    \caption{Methodology of measuring the planet radius gap in 3D. $Left$: $\log_{10}\left(\mathrm{Planet\ Radius}\right)$ versus $\log_{10}\left(\mathrm{Incident\ Flux}\right)$ at a mass slice of 1.00 $\mathrm{M_\odot}$. Contours represent a slice of the 3D KDE, while the dots represent planets re-drawn from the contours. We use \texttt{gapfit} \citep{Loyd2020} and 1000 bootstrap simulations with replacement to determine the best-fit gap line, which is shown in solid red, with the 16th and 84th percentile ranges for each incident flux value as the shaded region above and below the line. The gap's slope ($\alpha$) and intercept ($\log_{10}\left(R_{\mathrm{p,100}}\right)$) and their uncertainties for this individual mass slice are displayed at the bottom of the plot. $Right$: Best-fit line (red) to the 1000 bootstrapped $\log_{10}\left(R_{\mathrm{p,100}}\right)$ values (black circles) computed for each mass slice as in the left panel, as a function of $\log_{10}\left(\mathrm{Stellar\ Mass}\right)$. We have also plotted the 16th and 84th percentile ranges of this best-fit line as the red shaded region. In the right panel, we report $\alpha$ as the median and 16th and 84th percentiles of the 1000 gap slopes determined for each of the 15 stellar mass slices and $\beta$ as the slope (and 16th and 84th percentiles) of the red line in the right panel.}
    \label{fig:methods}
\end{figure*}

To measure the planet radius gap in 3D, we begin with the 3702 \kep\ confirmed and candidate planets between 0.7--5.1 \rearth\ and 0.1--10$^4$ \fearth\ with uncertainties smaller than their measured parameters and compute a 3D kernel density estimate (KDE) distribution in $\log_{10}\left(\mathrm{Planet\ Radius}\right)$, $\log_{10}\left(\mathrm{Incident\ Flux}\right)$, and $\log_{10}\left(\mathrm{Stellar\ Mass}\right)$ using \texttt{fastKDE} \citep{Obrien2014,Obrien2016}. We chose these bounds to contain roughly 95\% of the observed planet population smaller than 7 \rearth, center our analysis on super-Earths and sub-Neptunes, and maximize the KDE's resolution of the planet radius gap without ignoring significant portions of the super-Earths and sub-Neptunes. Next, we follow the procedure of \citet{Rogers2021b} by computing the mean stellar host mass ($\approx$ 0.99 \msun) and host mass standard deviation ($\approx$ 0.23 \msun) of the 3702 planets, and then split the 3D KDE data into 15 stellar mass slices, ranging in equal logarithmic steps from 0.76--1.22 \msun.

For each of these mass slices in planet radius versus incident flux-space, we used \texttt{gapfit} \citep{Loyd2020}, which takes discrete data in any two-dimensional parameter space, computes a kernel density estimate along a line that separates the ``gappy'' data, minimizes the density along that line by varying the line's slope and intercept, and then bootstraps with replacement \texttt{n} times to determine uncertainties in the best-fit line. We used the following initialization parameters:  \texttt{x0}=2.0 (reference $\log_{10}\left(\mathrm{Incident\ Flux}\right)$ in the gap line equation), \texttt{y0\_guess}=0.25 (an initial guess of the y-value of the gap at \texttt{x0}), \texttt{m\_guess}=0.05 (an initial guess of the slope of the gap), \texttt{sig}=0.15 (kernel width for \texttt{gapfit} KDE), \texttt{y0\_rng}=0.20 (the maximum by which the gap lines are allowed to deviate from \texttt{y0}), and \texttt{n}=1000 (number of bootstrap simulations with replacement). Because \texttt{gapfit} requires a discrete planet population, we drew 3702 planets in planet radius-incident flux-space based on the likelihood defined by our contours. The left panel of Figure \ref{fig:methods} illustrates this procedure at an individual mass slice of 1 \msun.

We repeated our \texttt{gapfit} procedure for each stellar mass slice, recording 1000 bootstrapped $\alpha_i$ values and intercept values ($\log R_{\mathrm{p,100},i}$), defined as the planet radius intersected by the best-fit gap line at 100 \fearth, for each slice. Next, we plotted the intercept values for each slice as a function of $\log_{10}\left(\mathrm{Stellar\ Mass}\right)$, and then fit 1000 lines to this data, one for each bootstrap simulation. We then computed the median of the slopes of these 1000 lines to determine $\beta$ $\equiv$ $\left(\partial \log R_{\mathrm{gap}} / \partial \log M_\star \right)_{S}$ and used the 16th and 84th percentiles of the best-fit lines to determine typical uncertainties on $\beta$. We show this fit in the right plot of Figure \ref{fig:methods}.

To determine the full uncertainties on $\alpha$ and $\beta$, we ran 1000 separate Monte Carlo simulations of the observed population, drawing each planet's radius, incident flux, and stellar mass values from a probability distribution defined by its radius, incident flux, and stellar mass and their corresponding uncertainties. We only kept simulated planets within the 0.7--5.1 \rearth\ and 0.1--10$^4$ \fearth\ bounds, which resulted in 10--50 fewer planets per simulation than the 3702 observed planets. We chose these bounds to match our observed population analysis above, center the super-Earths and sub-Neptunes, and maximize the KDE's resolution of the planet radius gap without ignoring significant portions of the simulated super-Earths and sub-Neptunes. We measured $\alpha$ and $\beta$ following the same procedure as above and as illustrated in Figure \ref{fig:methods}, recording 15000 unique $\alpha$ and 1000 unique $\beta$ values for each Monte Carlo simulation. Finally, we combined the 15 million $\alpha$ and 1 million unique $\beta$ values and computed the 16th, 50th, and 84th percentiles for each.

\section{Results \& Discussion}
\label{sec:results}

\begin{figure*}
    \centering
    \resizebox{\hsize}{!}{\includegraphics{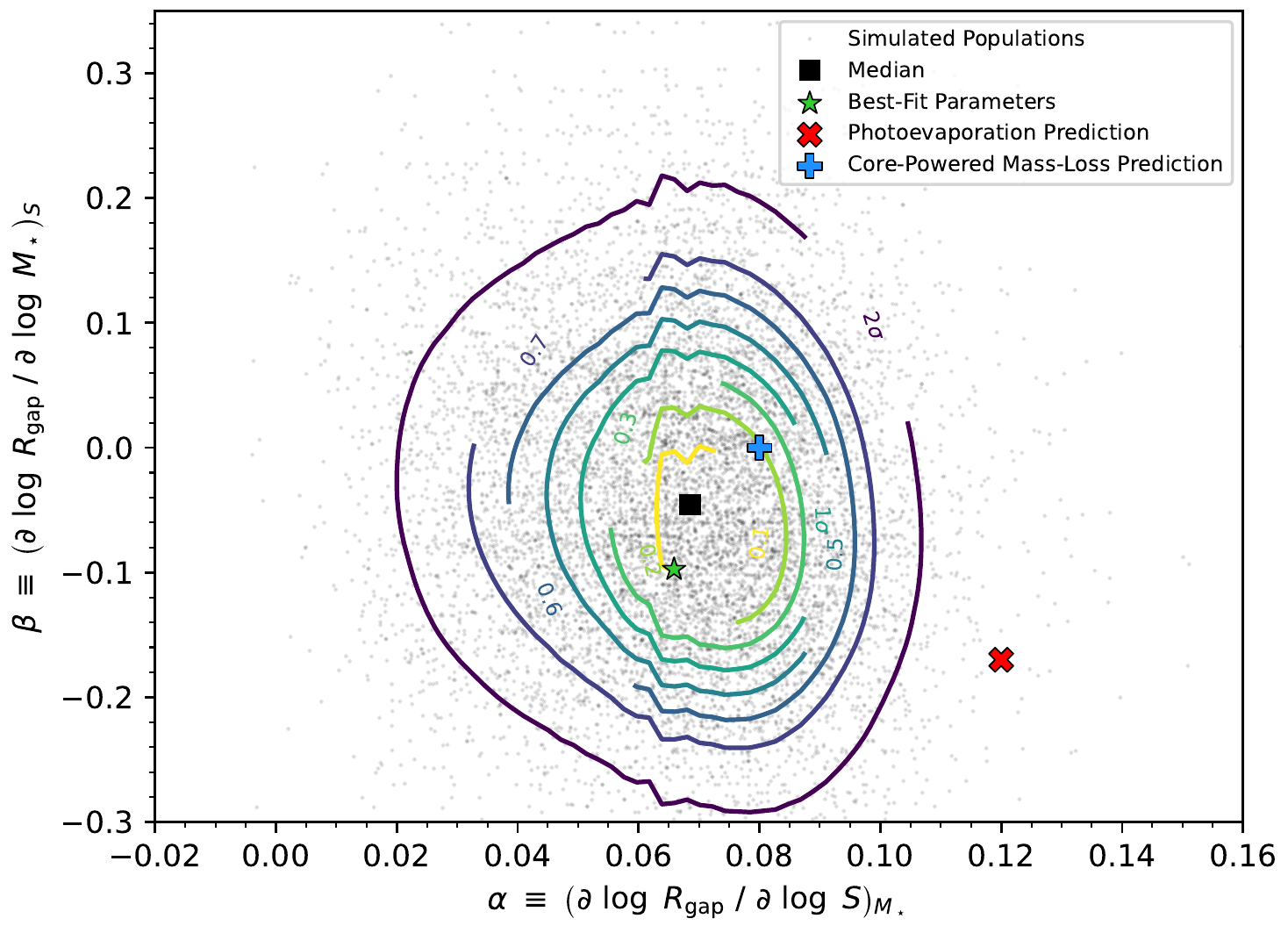}}
    \caption{$\beta$ versus $\alpha$ for the \kep\ planet population. The green star represents the observed distribution, while the filled red X and the filled blue plus represent the estimates of photoevaporation and core-powered mass-loss from \citep{Rogers2021b}, respectively. The grey dots show the gap measurements resulting from 1000 Monte Carlo simulations with 15 stellar mass slices per simulation and 1000 bootstraps per mass slice for a total of 15 million $\alpha$ measurements and 1 million unique $\beta$ measurements. The contours display a Kernel Density Estimate (KDE) of the resulting probability distribution of the Monte Carlo simulations which account for the effects of finite sampling. Contours are labeled according to the fraction of simulations contained within them, with the 1$\sigma$ and 2$\sigma$ contours defined as containing 39.35\% and 86.47\% of the simulations, respectively \citep{WangB2015}. The black square is the median result of the simulations.}
    \label{fig:3DGap}
\end{figure*}

We plot our results in Figure \ref{fig:3DGap}. We measure $\alpha$ = \calcalpha\ and $\beta$ = \calcbeta. The best-fit parameters occur within the contours representing the Monte Carlo simulations. The contours display higher variation on the $\beta$ axis than the $\alpha$ axis with a predominantly vertical orientation. The best-fit and median $\alpha$ values are smaller than predicted by both core-powered mass-loss and photoevaporation, while the best-fit and median $\beta$ values occur between the $\beta$ values predicted by core-powered mass-loss and photoevaporation. In addition, we find that the best-fit parameters are not centered within the contours, although they are well-within the uncertainties on both $\alpha$ and $\beta$ and can be explained by the random variations between simulations.

Given that the core-powered mass-loss prediction occurs at a contour corresponding to $\approx$ 0.7$\sigma$ and photoevaporation occurs at a contour corresponding to $\approx$ 2.8$\sigma$, we infer that core-powered mass-loss may be the dominant mechanism shaping the observed \kep\ planet population, assuming our methodology is not biased (e.g. see discussion below). Another potential complication is that photoevaporation's predicted $\beta$ is defined as the slope of the {\it upper edge} of the super-Earth population as a function of stellar mass at constant incident flux. We are unable to measure this quantity directly, as our measurement finds where the gap is {\it deepest}. Given the difference between a gap defined by the upper edge of the super-Earth population and where it is deepest, \citet{Rogers2021b} illustrates that the {\it deepest gap line} produces a {\it less-negative} $\beta$ than the super-Earth {\it upper edge gap line} that photoevaporation predicts. However, even if we were to shift the photoevaporation estimate (red filled X) up to a larger $\beta$, it still would not occur at a lower sigma contour than core-powered mass-loss.

To ensure our simulations were not biased relative to the observed population, we ran similar simulations as in \S \ref{sec:methods} but now reduced our uncertainties by factors of two and five. The resulting contours converged on the best-fit $\alpha$ and $\beta$ values with larger reductions in uncertainties, as expected. We also explored the effect of averaging the simulations together and computed contours that were virtually identical to the observed population. Therefore, we conclude that any biases between the simulated and best-fit parameters are introduced through the computation of the KDE from each simulated population and/or the subsequent determination of the best-fit gap line. However, these effects are small ($\Delta_\alpha$ = 0.004, $\Delta_\beta$ = 0.025) and well-within our reported uncertainties of $\sigma_\alpha$ $\approx$ 0.023 and $\sigma_\beta$ $\approx$ 0.130. As in \citet{Rogers2021b}, we wanted to ensure that we were using a sufficient number of stellar mass bins for computing $\alpha$ and $\beta$. Therefore, we tested 50 stellar mass bins in addition to the adopted 15 above and found no meaningful impact on our results.

We also performed the \S\ref{sec:methods} procedure for the \citet{Berger2020b} \kep\ planet sample to confirm our result is not sensitive to the exact stellar/planet properties. We computed $\alpha$ = 0.068$^{+0.018}_{-0.023}$ and $\beta$ = 0.042$^{+0.111}_{-0.109}$, which are statistically consistent with the $\alpha$ and $\beta$ values reported in this paper. In addition, we find that the predictions of core-powered mass-loss (corresponding contour $\approx$ 0.6$\sigma$) and photoevaporation (corresponding contour $\approx$ 3.1$\sigma$) for the \citet{Berger2020b} \kep\ planet sample are at similar likelihoods as those studied here.

To determine whether our methodology introduced any biasing systematics, we tested a variety of gap determination methods and found some discrepancies. We used four different gap determination methods, detailed below, which differ solely by their determination of the best-fit gap line and the corresponding slope ($\alpha_i$) as a function of incident flux and intercept ($\log R_{\mathrm{p,100},i}$) for each mass slice. Once each method has determined its $\alpha_i$ and $\log R_{\mathrm{p,100},i}$ values for each mass slice $i$, the determination of $\alpha$ and $\beta$ follows from the description in \S\ref{sec:methods}.
\begin{enumerate}
    \item The \citet{Petigura2022} methodology, which (1) draws a number of vertical lines spanning the gap as a function of incident flux, (2) finds the relative minimum density along those lines, and (3) fits a line to those minimum density points along each line.
    \item A modified \citet{Petigura2022} methodology, where instead of drawing vertical lines, we draw a line between the peaks of the super-Earth and sub-Neptune populations and then draw corresponding parallel lines which span the gap. We use the same density arguments as in method 1 to determine the points on each line to fit.
    \item The \citet{Rogers2021b} methodology, which relies on minimizing the sum of the KDE along the gap line.
    \item Our \texttt{gapfit} methodology, which we detailed in \S\ref{sec:methods} above.
\end{enumerate}
We found that while methodologies 1 and 3 produced similar results and 2 and 4 produced similar results, comparisons between the pairs produced systematically different results. Interestingly, we found systematically larger $\alpha$ values using the \emph{vertical} lines of \citet{Petigura2022} than if we had repeated the same procedure with \emph{diagonal} lines connecting the sub-Neptune and super-Earth peaks. This suggests that the methods dependent on drawing lines to define the gap as a function of incident flux are inherently dependent on the exact orientation of the lines drawn. In addition, the \citet{Rogers2021b} method (method 3) relies on minimizing the sum of the KDE along the gap line, although we found that the method failed occasionally depending on the exact simulated planet population. Upon inspection, both gap line orientation methods appeared to fit the gap equally well, so we concluded that we needed a method that would capture this uncertainty.

Hence, we decided to use \texttt{gapfit} \citep{Loyd2020}, which, similar to \citet{Rogers2021b}, attempts to minimize the sum of the KDE along the gap line and also bootstraps with replacement to sample the gap line uncertainties for each mass slice of each Monte Carlo-simulated planet population. We also found that our method is robust and rarely failed to produce $\alpha$ and $\beta$ estimates after hundreds of simulations. Regardless, we produce systematically smaller $\alpha$ values than those measured by \citet{Rogers2021b}, which results in our $\gtrsim$2$\sigma$ preference for core-powered mass-loss over photoevaporation as the mechanism sculpting the \kep\ exoplanet radius gap. Interestingly, \citet{Ho2023} used \kep\ short cadence observations and yet another independent gap fitting approach to find $\alpha$ and $\beta$ values that are more consistent with core-powered mass-loss than photoevaporation.

We note that the $expectation$ values of $\alpha$ and $\beta$ for the two theories determined by \citet{Rogers2021b}, to which we compare our results, are not the same as those $measured$ by \citet{Rogers2021b}. However, we caution that any measurement of the 3D planet radius gap may be the result of systematics present in the chosen gap-determination method, given the discrepancies between our measurements and those of \citet{Rogers2021b} and \citet{Petigura2022}. This therefore motivates the use of an unbiased gap measurement methodology and a need for more precise and accurate planet radii.

\section{Conclusion}
\label{sec:conclusion}

In this paper, we used the latest \kep\ host stellar properties with \gaia\ DR3 constraints from \ref{B23} to compute the slope of the \kep\ planet radius gap as a function of incident flux and stellar mass. We aimed to differentiate between the mechanisms of core-powered mass-loss and photoevaporation for \kep\ planets.

We measure $\alpha$ = \calcalpha\ and $\beta$ = \calcbeta, which are $\gtrsim$2$\sigma$ more consistent with the \citet{Rogers2021b} predictions of core-powered mass-loss ($\alpha$ $\approx$ 0.08, $\beta$ $\approx$ 0.00) than photoevaporation ($\alpha$ $\approx$ 0.12, $\beta$ $\approx$ --0.17) and consistent with \citet{Ho2023}. While photoevaporation predicts a more-negative $\beta$ than is possible to measure due to photoevaporation's gap definition as the upper edge of the super-Earth population rather than where the gap is deepest, shifting the photoevaporation prediction to a larger $\beta$ does not change our conclusion.

We caution that the measurement of the exoplanet radius gap is dependent on the exact methodology used. We found that our methodology based on \texttt{gapfit} \citep{Loyd2020} produced systematically smaller $\alpha$ values relative to the \citet{Rogers2021b} and \citet{Petigura2022} methods, although they are consistent within uncertainties. \citet{Rogers2021b} measures an $\alpha$ = 0.10$^{+0.03}_{-0.02}$ that is consistent with the theoretical predictions of both core-powered mass-loss ($\alpha$ $\approx$ 0.08) and photoevaporation ($\alpha$ $\approx$ 0.12), while our systematically smaller $\alpha$ = \calcalpha\ favors core-powered mass-loss over photoevaporation. Our measured $\beta$ = \calcbeta\ is consistent with both theories.

While adding more planets from \ktwo\ and \tess\ may improve our understanding of planet demographics, the large number and exquisite precision of \kep\ planet radii represent the ideal sample for constraining $\alpha$ and $\beta$ and hence the planet radius gap. This study therefore motivates a systematic re-fitting of \kep\ planet transits, leveraging the latest eccentricity priors for single and multi-planet systems and mean stellar densities to better constrain \rprstar\ values \citep{Petigura2020}. In turn, this will increase the accuracy and precision of planet radii and hence our measurements of the planet radius gap. These new measurements may enable us to differentiate between the theories of core-powered mass-loss and photoevaporation for the \kep\ sample once and for all.

\software{\texttt{fastKDE} \citep{Obrien2014,Obrien2016}, \texttt{gapfit} \citep{Loyd2020}, \texttt{isoclassify} \citep{Huber2017,Berger2020a} \texttt{kiauhoku} \citep{Claytor2020}, \texttt{matplotlib} \citep{Matplotlib}, \texttt{numpy} \citep{numpy}, \texttt{pandas} \citep{Pandas}, \texttt{scipy} \citep{Scipy}, \texttt{TOPCAT} \citep{topcat}}

\facility{Exoplanet Archive}

\begin{acknowledgments}
We thank James Rogers, Akash Gupta, James Owen, Hilke Schlichting, Vincent Van Eylen, and Eric Gaidos for their helpful discussions regarding the results and methodology. This research has made use of the NASA Exoplanet Archive, which is operated by the California Institute of Technology, under contract with the National Aeronautics and Space Administration under the Exoplanet Exploration Program. This work has made use of data from the European Space Agency (ESA) mission {\it Gaia} (\url{https://www.cosmos.esa.int/gaia}), processed by the {\it Gaia} Data Processing and Analysis Consortium (DPAC, \url{https://www.cosmos.esa.int/web/gaia/dpac/consortium}). Funding for the DPAC has been provided by national institutions, in particular the institutions
participating in the {\it Gaia} Multilateral Agreement. T.A.B.’s research was supported by an appointment to the NASA Postdoctoral Program at the NASA Goddard Space Flight Center, administered by Universities Space Research Association and Oak Ridge Associated Universities under contract with NASA. D.H. acknowledges support from the Alfred P. Sloan Foundation and the National Aeronautics and Space Administration (80NSSC19K0597, 80NSSC21K0652).
\end{acknowledgments}

\bibliography{3DGap.bib}{}
\bibliographystyle{aasjournal}


\end{document}